\documentclass[aps,prc,twocolumn,amsmath,amssymb,floatfix,superscriptaddress]
{revtex4-1}
\usepackage{silence} 
\usepackage{mathptmx}
\usepackage[T1]{fontenc}
\usepackage{CJKutf8}
\usepackage{graphicx}
\usepackage{bm}
\usepackage{dcolumn}
\usepackage{multirow}
\usepackage{xcolor}
\usepackage{booktabs}
\usepackage{comment}
\usepackage{soul}
\usepackage[
breaklinks,pdfstartview=FitH,CJKbookmarks=true,
bookmarksnumbered=true,bookmarksopen=true,pdfborder={0 0 1},
colorlinks=true,linkcolor=blue,urlcolor=blue,anchorcolor=blue,citecolor=blue
            ]{hyperref}

\def\be{\begin{equation}} \def\ee{\end{equation}}
\def\bal#1\eal{\begin{align}#1\end{align}}
\def\non{\nonumber}
\long\def\hj#1{\color{red}#1\color{black}}
\long\def\zx#1{\color{blue}#1\color{black}}
\long\def\OFF#1{}

\def\eps{\varepsilon}
\def\la{\Lambda}
\def\rv{\bm{r}} \def\kv{\bm{k}} \def\Jv{\bm{J}} \def\Wv{\bm{W}} \def\Nv{\bm{\nabla}}
\def\ron{\rho_N} \def\tn{\tau_N}
\def\rol{\rho_\la} \def\tl{\tau_\la} \def\ml{m_\la}

\def\fmq{\;\text{fm}^{-3}}
\def\mev{\;\text{MeV}}
\def\mfmq{\;\text{MeV}\,\text{fm}^{-3}}
\def\pfmq{\;\text{MeV}\,\text{fm}^3}
\def\lpb{$^{208}_{\hskip0.6em\la}$Pb}
\def\ltl{$^{208}_{\hskip0.7em\la}$Tl}
\def\lla{$^{139}_{\hskip0.7em\la}$La}
\def\lys{$^{89}_{\ \la}$Y}
\def\lv{$^{51}_{\ \la}$V}
\def\lsi{$^{28}_{\ \la}$Si}
\def\lks{$^{40}_{\ \la}$K}
\def\lka{$^{48}_{\ \la}$K}

\def\lo{$^{16}_{\ \la}$O}

\def\lcl{$^{13}_{\ \la}$C}

\def\renv{\langle\rho\rangle_\text{env}}
\def\renv{\bar{\rho}_\text{env}}

\begin{document}

\begin{CJK*}{UTF8}{gbsn}
\title{
Density dependence of
effective $\pmb{\la}$-nucleon interactions
}

\author{Xuan Zeng (曾炫)}
\affiliation{
Department of Physics, East China Normal University,
Shanghai 200241, China}

\author{Xian-Rong Zhou (周先荣)} \email{xrzhou@phy.ecnu.edu.cn}
\affiliation{
Department of Physics, East China Normal University,
Shanghai 200241, China}

\author{Yu-Gang Ma (马余刚)} \email{mayugang@fudan.edu.cn}
\affiliation{
Department of Physics, East China Normal University,
Shanghai 200241, China}
\affiliation{
Key Laboratory of Nuclear Physics and Ion-beam Application (MOE),
Institute of Modern Physics, Fudan University, Shanghai, 200433, China}
\affiliation{
Shanghai Research Center for Theoretical Nuclear Physics,
NSFC and Fudan University, Shanghai, 200438, China}

\author{H.-J. Schulze}
\affiliation{
INFN Sezione di Catania, Dipartimento di Fisica, Universit\`a di Catania,
Via Santa Sofia 64, 95123 Catania, Italy}

\date{\today}

\begin{abstract}
Orbital-dependent $\la$ binding energies in hypernuclei
such as \lpb, \lks, and \lka\
are important objectives of planned experiments at J-PARC and JLab
and provide information on the density-dependent features
of effective $\la N$ interactions.
To investigate this relation we introduce the environmental density $\renv$,
defined as the overlap integral of the $\la$ and nuclear core densities,
as quantitative measure of the nuclear environment
experienced by the $\la$ in each orbital.
Calculations are performed
within the deformed Skyrme-Hartree-Fock framework,
comparing several phenomenological and microscopic $\la N$ interactions.
We find an approximately linear relation
between the orbital-dependent binding energies and $\renv$,
offering a quantitative means to test and refine effective $\la N$ interactions
using future orbital-dependent binding-energy data.
\end{abstract}

\maketitle

\end{CJK*}

\section{Introduction}
\label{intro}

A hyperon carries a strangeness quantum number that distinguishes it
from ordinary nucleons,
and is therefore not constrained by the Pauli exclusion principle,
making it a flavor-tagged internal probe for investigating nuclear properties
\cite{Bydzovsky07,Gal16,Vidana21,Ma23}.
In addition,
several anti-matter hypernuclei have been observed
at ultra-relativistic heavy-ion colliders
\cite{Abelev10,Abdulhamid24,Acharya25,Chen25}.
With the advancement of experimental techniques,
high-precision measurements of single-$\la$ binding energies
can now be achieved over a broad mass range,
from light to heavy hypernuclei
\cite{Hashimoto06,Feliciello15,Gal16,Adam20,Abdallah22,Kino26}.
Therefore, the hyperon-nucleon ($YN$) interaction,
which incorporates the strangeness degree of freedom,
is a key to understanding various novel experimental results on hypernuclei
and can provide further insights into the strong interaction
\cite{Davis05,Hashimoto06,Fujiwara07,Schulze10,Feliciello15,Friedman23,Zheng25}.

The nucleon-nucleon ($NN$) interaction is tightly constrained
by abundant $NN$ scattering data
\cite{Stoks94,Wiringa95,Machleidt01,Epelbaum15,Entem17},
whereas the $YN$ interaction is much less constrained
because the available $YN$ scattering data are sparse
\cite{Engelmann66,Cline68,Sechi68,Alexander68,Eisele71,Kadyk71,Hauptman77}.
We mention that recent advances in nuclear femtoscopy have been
able to reveal complementary information
on low-energy $YN$ scattering parameters
inaccessible to scattering experiments
\cite{Verde06,Fabbietti21,Adamk19,Acharya20,Acharya20b,Acharya24,Yan25}.

However,
experimental observations of finite hypernuclei offer useful
complementary information to constrain the $YN$ interaction.
For example,
systematic measurements of the binding of the $\la$ hyperon
in different shells of hypernuclei
provide valuable density-dependent information on the $\la N$ interaction
\cite{Schulze10,Isaka16,Gal16},
while observations of mirror $\la$ hypernuclei
impose additional constraints on its charge-symmetry-breaking (CSB) properties
\cite{Juric73,Nakamura13,Gal15,Gazda16,Sun25}.
Observations of neutron stars with masses around $2M_\odot$
constrain the equation of state of hypernuclear matter
and the high-density behavior of hyperonic interactions
\cite{Schulze11,Colucci13,Hu14,Fonseca16,Haidenbauer17,
Tolos20,Zheng25,Vidana13}.

Therefore,
given the limited direct constraints from $YN$ scattering and femtoscopy,
ongoing and future hypernuclear experiments,
together with neutron-star observables,
provide essential tests of effective $\la N$ interactions
and their density dependence.
Heavy hypernuclei are particularly useful because a $\la$ can occupy orbitals
that sample different regions of the same nuclear core
and thus probe the density dependence of the $\la N$ interaction.
In particular, the currently heaviest hypernucleus \lpb\
allows the $\la$ hyperon to occupy orbitals ranging from $s$ to $g$,
making it an excellent system for probing the density dependence
of the $\la N$ interaction.
Future high-resolution measurements of these states
at the HIHR beamline of J-PARC are therefore of special interest
\cite{Nakamura22}.
The planned measurements of the neutron-rich heavy hypernucleus \ltl\
and the lighter hypernuclei \lks\ and \lka\ at JLab
will provide complementary information over a broad range of mass
and isospin asymmetry \cite{Nakamura22,Gogami24}.

In the theory of hypernuclei,
the deformed Skyrme-Hartree-Fock (DSHF) approach
is an efficient mean-field method
that is not hindered by the increasing mass of the nuclear core
\cite{Vautherin72,Vautherin73,Bender03,Stone07,Erler11}.
It is applicable to nuclei across a wide range of mass numbers
and enables global calculations over the entire nuclide chart
for $\la$ hypernuclei
\cite{Zhou07,ChenCF22,ChenYF22,Guo22,Xue22,Xue23,Xue24}.
Furthermore, through angular-momentum projection
and the generator-coordinate method,
DSHF can be extended beyond the mean-field approximation
to provide a comprehensive description of the energy spectra of hypernuclei
\cite{Cui15,Cui17,Cui17b,Xue24}.

Within this approach,
$\la N$ interactions can be classified into Skyrme-type and microscopic-type.
Skyrme-type $\la N$ interactions are constructed by fitting empirical parameters
to hypernuclear experimental data,
yielding different parameter sets depending on the specific research goal
\cite{Rayet76,Rayet81,Yamamoto88,Yamamoto10,Schulze19,Xue23,Qiu25,Liu25,Liu26}.
They can efficiently yield favorable theoretical results,
among which the latest SLL4 interaction
\cite{Schulze14,Schulze19}
has achieved a certain degree of success.
RAY12 is an earlier Skyrme-type parametrization developed mainly from
the available data on light hypernuclei \cite{Rayet81}.
It contains no effective-mass correction,
corresponding to $m_\la^*/m_\la=1$,
and thus provides a useful reference for examining
how effective-mass differences affect the orbital behavior of the $\la$ hyperon.

However, the parameters obtained from fitting often lack
a deeper microscopic theoretical foundation.
Microscopic interactions are often based on the $G$ matrix calculated
using the Brueckner-Hartree-Fock (BHF)
or relativistic BHF (RBHF) method,
and are combined with the SHF approach
\cite{Cugnon00,Yamamoto10,Vidana01,Zhou07,Schulze08,Schulze10,Schulze11,Zeng26}.
Specifically, the bare nuclear force is first extended
to an in-medium force using the (R)BHF method,
which is then adapted to finite nuclear systems
by a local-density approximation
through its single-particle (s.p.) potential in nuclear matter.
The comparison between theoretical predictions and experimental observations
often reveals more microscopic theoretical information
about the bare nuclear interaction.
Since the NSC89 $YN$ interaction \cite{Cugnon00}
derived from the meson-exchange Nijmegen potentials \cite{Maessen89},
various microscopic interactions
have been successively extended to finite hypernuclear systems,
such as NSC97f \cite{Rijken99,Vidana01},
ESC08 \cite{Schulze13},
and RCh-LO700 \cite{Zheng25,Zeng26},
the latter derived from a relativistic chiral effective interaction,
leading to promising predictive results
\cite{Cugnon00,Vidana01,Schulze11,Schulze13}.
However, compared with the directly fitted phenomenological interactions,
the microscopic interactions are slightly less accurate in describing the data.

Regardless of their phenomenological or microscopic origin,
the density-dependent properties of effective $\la N$ interactions are essential
for describing observables in hypernuclei.
They affect the binding energies of the $\la$ hyperon in different orbitals,
which can be directly compared with experimental results.
They also influence the spatial distribution of the $\la$ wave function,
and such changes can in turn affect limiting structural properties
of hypernuclei \cite{Zhou08,Zhang21,Ding23,Liu25}.

Although the density-dependent characteristics of effective $\la N$
interactions have been qualitatively examined using different theoretical
approaches \cite{Isaka16,ChenYF22,Ding23},
establishing a quantitative connection with the orbital-dependent $B_\la$
requires a characteristic measure of the nuclear environment
experienced by the $\la$ in each orbital.
This may help to make the differences among the interactions more transparent
and to assess how they may be further tested and refined.
This is particularly relevant for the recently developed
microscopic interaction RCh-LO700,
whose density-dependent behavior in finite hypernuclei has not yet been
systematically examined.

Motivated by these considerations,
we introduce the environmental density $\renv$
and evaluate it for selected hypernuclei,
including \lpb, \lks, and \lka,
which are of particular interest to planned experiments at J-PARC and JLab.
Calculations are performed within the DSHF framework including
Bardeen-Cooper-Schrieffer (BCS) pairing correlations
\cite{Tajima93,Shen99,Bender00,Suzuki03,Ring04,Zhou07},
using the phenomenological interactions RAY12 and SLL4
and the microscopic interactions RCh-LO700 and NSC97f.
The analysis provides a framework for quantitatively comparing,
testing, and refining effective $\la N$ interactions
as new orbital-dependent binding-energy data become available.

The paper is organized as follows.
In Sec.~\ref{s:form},
a brief introduction to the theoretical DSHF approach for hypernuclei is given,
along with the specific forms of the Skyrme-type and microscopic interactions
within the SHF framework.
In Sec.~\ref{s:resu},
theoretical predictions for the selected hypernuclei are presented,
and the environmental density is used to quantify and compare
the effective density dependence of the four interactions.
The implications for testing and refining these interactions,
particularly RCh-LO700,
are also discussed.
Finally, a summary is given in Sec.~\ref{s:conc}.

\section{Theoretical framework}
\label{s:form}

In the DSHF approach,
the total energy of a hypernucleus is given by
\cite{Rayet81,Cugnon00,Vidana01,Zhou07,Schulze08,Guo22,Xue24}
\be
 E = \int d^3\rv\, \eps(\rv) \:,
\label{e:EE}
\ee
where the energy-density functional is
\be
 \eps = \eps_N[\tau_n,\tau_p,\rho_n,\rho_p,\Jv_n,\Jv_p]
 + \eps_\la[\tau_\la,\rol,\ron,\Jv_N,\Jv_\la] \:,
\ee
where $\eps_N$ and $\eps_\la$ denote the energy-density functionals
for the nucleon and $\la$ parts,
which arise from the $NN$ and $\la N$ interactions, respectively.
They depend on the one-body density $\rho_q$,
kinetic-energy density $\tau_q$,
and spin-orbit current $\Jv_q$,
with $q = n, p, \la$ denoting the particle species.
These quantities are given by
\be
 \big[ \rho_q, \tau_q, \Jv_q \big]
 = \sum_{k=1}^{N_q} n_q^k \Big[
 \big| \phi_q^k \big|^2,
 \big| \Nv\phi_q^k \big|^2,
 {\phi_q^k}^* \big(\Nv\phi_q^k \times \bm{\sigma}\big)/i \Big] \:.
\ee
The s.p.~wave functions $\phi_q^k$ represent
the $k$th occupied states of different particle species $q$,
obtained through self-consistent calculations.
The occupation probabilities $n_q^k$ are obtained using the BCS approximation,
which is applied only to nucleons,
with the pairing interaction between nucleons described by a
density-dependent $\delta$ force \cite{Tajima93,Bender00},
\be
 V_q(\rv_1,\rv_2) = V_q' \bigg[ 1 -
 \frac{\ron\left((\rv_1+\rv_2)/2\right)}{0.16\,\text{fm}^{-3}} \bigg]
 \delta(\rv_1-\rv_2) \:,
\ee
where pairing strengths
$V_p' = V_n' = -410\pfmq$
are used for light ($A<40$) nuclei \cite{Suzuki03},
while $V_p' = -1146\pfmq$ and $V_n' = -999\pfmq$
are adopted for medium-mass and heavy nuclei \cite{Zhou07}.
A smooth energy cutoff is employed in the BCS calculations \cite{Bender00}.
In the case of an odd nucleon number,
the orbit occupied by the unpaired nucleon is blocked \cite{Ring04}.
\OFF{
For the hypernuclei considered here,
the BCS pairing contributions are either small
or nearly interaction independent.
They therefore do not qualitatively affect the comparison
of the orbital-dependent $\la$ binding energies
among the different effective $\la N$ interactions.
}

For the nucleon energy-density functional $\eps_N$,
we adopt the SLy4 parameterization,
which provides a good description of the binding energies
of light and heavy nuclei
\cite{Chabanat98,Bender03,Stone07,Aytekin10,Erler11}.
Hyperon observables are to a good extent insensitive
to the choice of the $NN$ interaction
\cite{Zhou07,ChenYF22,Xue22}.

For the hyperon energy-density functional $\eps_\la$,
we employ the Skyrme-type $\la N$ interactions RAY12 \cite{Rayet81}
and SLL4 \cite{Schulze19},
as well as the microscopic interactions RCh-LO700 \cite{Zeng26}
and NSC97f \cite{Vidana01}.
The former can be written as
\bal
 \eps_\la =& \frac{\tl}{2m_\la}
 + a_0\rol\ron + a_1(\rol\tn + \ron\tl)
\non\\&
 - a_2(\rol\Delta\ron + \ron\Delta\rol)/2 + a_3\rol\ron^{1+\alpha}
\non\\&
 - a_4(\rol \Nv\cdot\Jv_N + \ron\Nv\cdot \Jv_\la)  
\label{e:sll}
\eal
with a $\la$ effective mass
\be
 \frac{\ml^*}{\ml}(\ron) = \frac{1}{1 + 2\ml a_1 \ron} \:,
\label{e:mfit}
\ee
whereas the latter are
\cite{Cugnon00,Vidana01,Schulze08,Schulze10,Zeng26}
\bal
 \eps_\la =& \frac{1}{2\ml} \tl + \eps_{N\la}(\ron,\rol)
\\&
 + \bigg( \frac{\ml}{\ml^*(\ron)} - 1 \bigg)
 \bigg( \frac{\tl}{2\ml} - \frac35 \frac{(3\pi^2)^{2/3}}{2\ml} \rol^{5/3} \bigg)
 \:,
\eal
where
\bal
 \eps_{N\la}(\ron,\rol) =&
 - \left(\eps_1 - \eps_2\ron + \eps_3\ron^2 \right) \ron\rol
\non\\&
 + \left(\eps_4 - \eps_5\ron + \eps_6\ron^2 \right) \ron\rol^{5/3}
\non\\&
 - \left(\eps_7 - \eps_8\rol + \eps_9\rol^2 \right) \rol^2 \:,
\label{e:nsc}
\eal
and the effective mass is described by a simple phenomenological expression,
\be
 \frac{\ml^*}{\ml}(\ron) = \mu_1 - \mu_2\ron + \mu_3\ron^2 - \mu_4\ron^3 \:.
\label{e:mu}
\ee
It is worth noting that in BHF
the effective mass reflects the momentum dependence
of the s.p.~kinetic energy as modified by the nuclear medium
\cite{Baldo00,Cugnon00,Vidana00},
whereas in RBHF it is incorporated within the Dirac equation
to account for the exchange of scalar mesons
\cite{Hu14,Zheng25}.

\begin{figure}[t]
\vspace{3mm}
\includegraphics[scale=0.38]{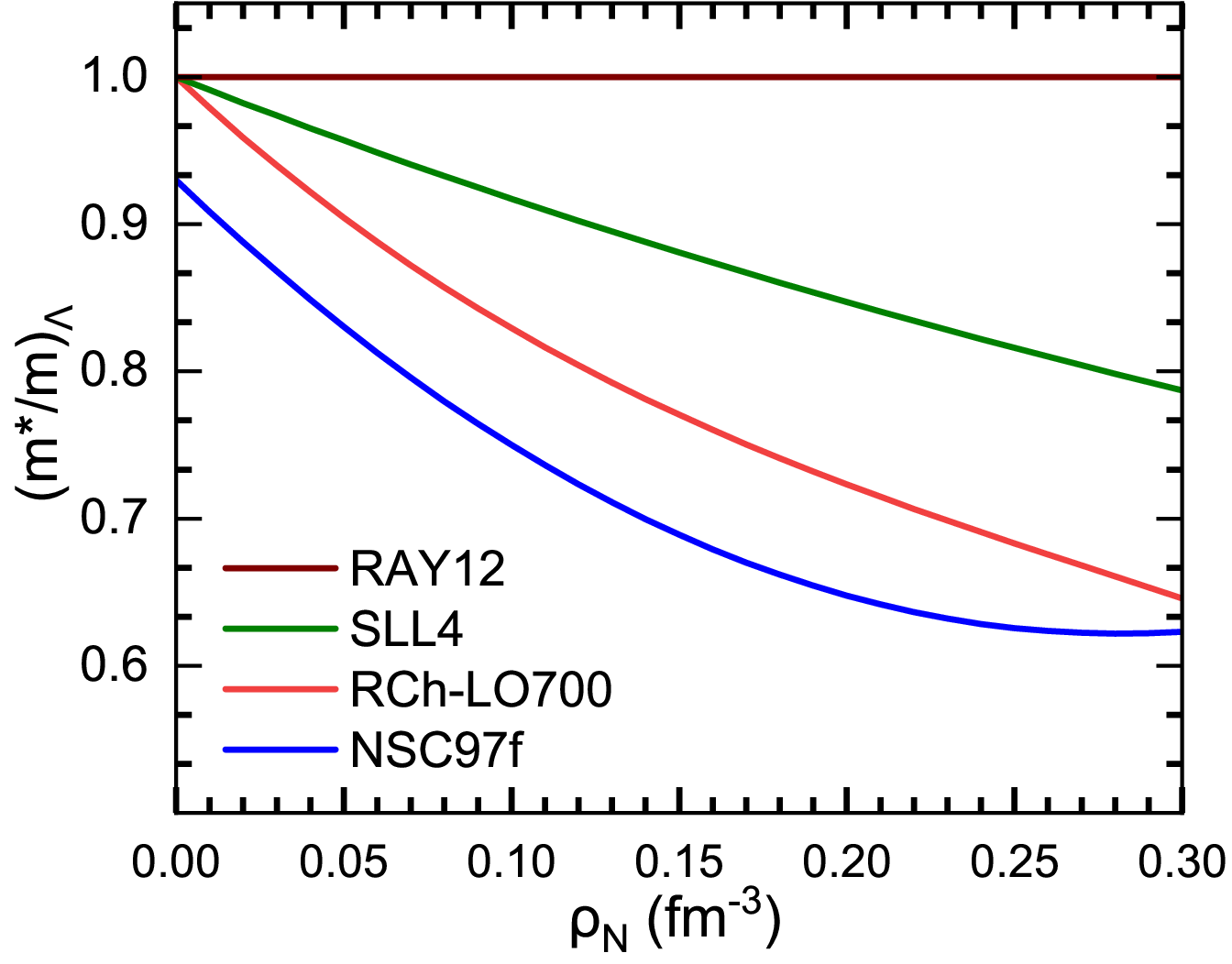}
\vspace{-6mm}
\caption{
Density dependence of the $\la$ effective mass in symmetric nuclear matter
for the different interactions.
}
\label{f:meff}
\end{figure}

Table~\ref{t:par} lists the parameters of the four interactions
under consideration,
consistent with densities $\ron$ and $\rol$ expressed in units of $\!\fmq$,
and $\eps_\la$ in units of $\!\mfmq$.
It is notable that the selected RAY12 force uses the bare $\la$ mass ($a_1=0$)
and a surface term ($a_2$) with opposite sign as SLL4.
For later use,
we present the density-dependent $\la$ effective masses
of the different interactions in Fig.~\ref{f:meff}.

\begin{table}[t]
\caption{
The parameters of the Skyrme-type interactions
RAY12 \cite{Rayet81} and SLL4 \cite{Schulze10},
Eq.~(\ref{e:sll}),
and the microscopic interactions
RCh-LO700 \cite{Zeng26} and NSC97f \cite{Vidana01},
Eqs.~(\ref{e:nsc}) and (\ref{e:mu}).
The values are consistent with
densities $\ron$ and $\rol$ expressed in units of $\!\fmq$,
and $\eps_\la$ in units of $\!\mfmq$.
}
\renewcommand\arraystretch{1.0}
\setlength{\tabcolsep}{9pt}
\def\mc#1{\multicolumn{2}{d}{#1}}
\begin{ruledtabular}
\begin{tabular}{lcc|rrr}
         &RAY12 & SLL4 &          & RCh-LO700 & NSC97f  \\
\midrule
 $a_0$    & -237.4  & -322.0 &$\eps_1$ & 369      & 384   \\
 $a_1$    & 0       & 15.75  &$\eps_2$ & 1402     & 1473  \\
 $a_2$    & -6.85   & 19.63  &$\eps_3$ & 1894     & 1933  \\
 $a_3$    & 864.75  & 715.0  &$\eps_4$ & 965      & 635   \\
$\alpha$ & 1       & 1      &$\eps_5$ & 4021     & 1829  \\
$a_4$    & 0       & 0      &$\eps_6$ & 4997     & 4100  \\
         &         &        &$\eps_7$ & 0        & 50    \\
         &         &        &$\eps_8$ & 0        & 545   \\
         &         &        &$\eps_9$ & 0        & 981   \\
         &         &        &$\mu_1$  & 1.00     & 0.93  \\
         &         &        &$\mu_2$  & 2.15     & 2.19  \\
         &         &        &$\mu_3$  & 5.03     & 3.89  \\
         &         &        &$\mu_4$  & 6.00     & 0.00  \\
\end{tabular}
\end{ruledtabular}
\label{t:par}
\end{table}

Through the variation of the total energy Eq.~(\ref{e:EE})
one derives the SHF Schr\"odinger equation for both nucleons and hyperons,
\be
 \bigg[ -\Nv \cdot \frac{1}{2m_q^*(\rv)} \Nv + V_q(\rv)
 - i\Wv_q(\rv) \cdot (\Nv\times \bm{\sigma}) \bigg] \phi_q^k(\rv)
 = e_q^k \phi_q^k(\rv) \:,
\ee
where $V_q$ is the density-dependent central part of the mean field
for different particles,
while $\Wv_q$ represents the spin-orbit interaction between nucleons
\cite{Vautherin73,Bender00}.
Since the $\la$ spin-orbit splitting is much smaller than its nucleonic
counterpart \cite{Ajimura01,Finelli09},
the corresponding term is neglected in this work ($a_4=0$).
The mean-field correction arising from the $\la$ is given by
the partial derivative of its corresponding energy density functional
with respect to the nucleon or hyperon density,
\bal
 V_N =& V_N^\text{SHF} + \frac{\partial\eps_\la}{\partial\ron} \:,
\\
 V_\la =& \frac{\partial\eps_\la}{\partial\rol} \:,
\label{e:potshf}
\eal
where $V^\text{SHF}_N$ denotes the Skyrme mean field for nucleons
in the absence of hyperons.

To account for the center-of-mass correction,
the standard substitution
\cite{Vautherin72,Vautherin73,Bender03,Stone07,Erler11}
\be
 \frac{1}{m_q} \rightarrow \frac{1}{m_q}-\frac{1}{M} 
\ee
is performed,
where $M = (N_n + N_p) m_N + N_\la m_\la$
is the total mass of the hypernucleus.

In our model approximation,
the mean field is assumed to be axially symmetric,
and the deformed SHF Schr\"odinger equation is solved in cylindrical coordinates
$(r,z)$ using a deformed harmonic oscillator basis
\cite{Vautherin73,Bender03,Stone07,Erler11}.
This allows us to calculate deformed hypernuclei under quadrupole constraints.
The optimal quadrupole deformation parameters
\be
 \beta_2^{(q)} \equiv \sqrt{\frac{\pi}{5}}
 \frac{\big\langle 2 z^2-r^2 \big\rangle_q}
 {\big\langle r^2+z^2 \big\rangle_q}
\ee
are determined by minimizing the total energy.

\section{Results and discussion}
\label{s:resu}

We employ the interactions RAY12, SLL4, \hbox{RCh-LO700}, and NSC97f
to perform systematic predictions for \lpb, \lks, \lka, and other hypernuclei,
focusing on the $\la$ separation energy
\be
 -B_\la = E(^A_{\la}Z) - E(^{A-1}Z) 
\ee
of the different $\la$ orbitals,
which depends on the deformation $\beta_2^{(N)}$ in the DSHF formalism.

The spin-orbit interaction is not taken into account
in any of the $\la N$ interactions.
Consequently, in the spherical limit,
the $j=l\pm1/2$ partners are degenerate.
In a deformed system, $\beta_2 \neq 0$,
$j$ is no longer a good quantum number:
deformation splits states according to their spatial Nilsson quantum numbers,
while the two spin projections of a given spatial state remain degenerate.
This behavior is consistent with Ref.~\cite{Xue23}.

Since \lpb\ and \ltl\ differ only very slightly in isospin
and our interactions barely include isospin effects,
the calculated results for both nuclei are nearly identical.
Therefore, we only discuss \lpb\ here.
Future precise measurements of the $\la$ binding energies
in different s.p.~orbitals of \lpb\ at HIHR
will greatly facilitate the discussion of the exact strengths
of the $\la N$ central and spin-orbit interactions and isospin effect
\cite{Sakuma22}.

\subsection{Binding energies}

\begin{figure}[t]
\vspace{-0mm}
\includegraphics[scale=0.42]{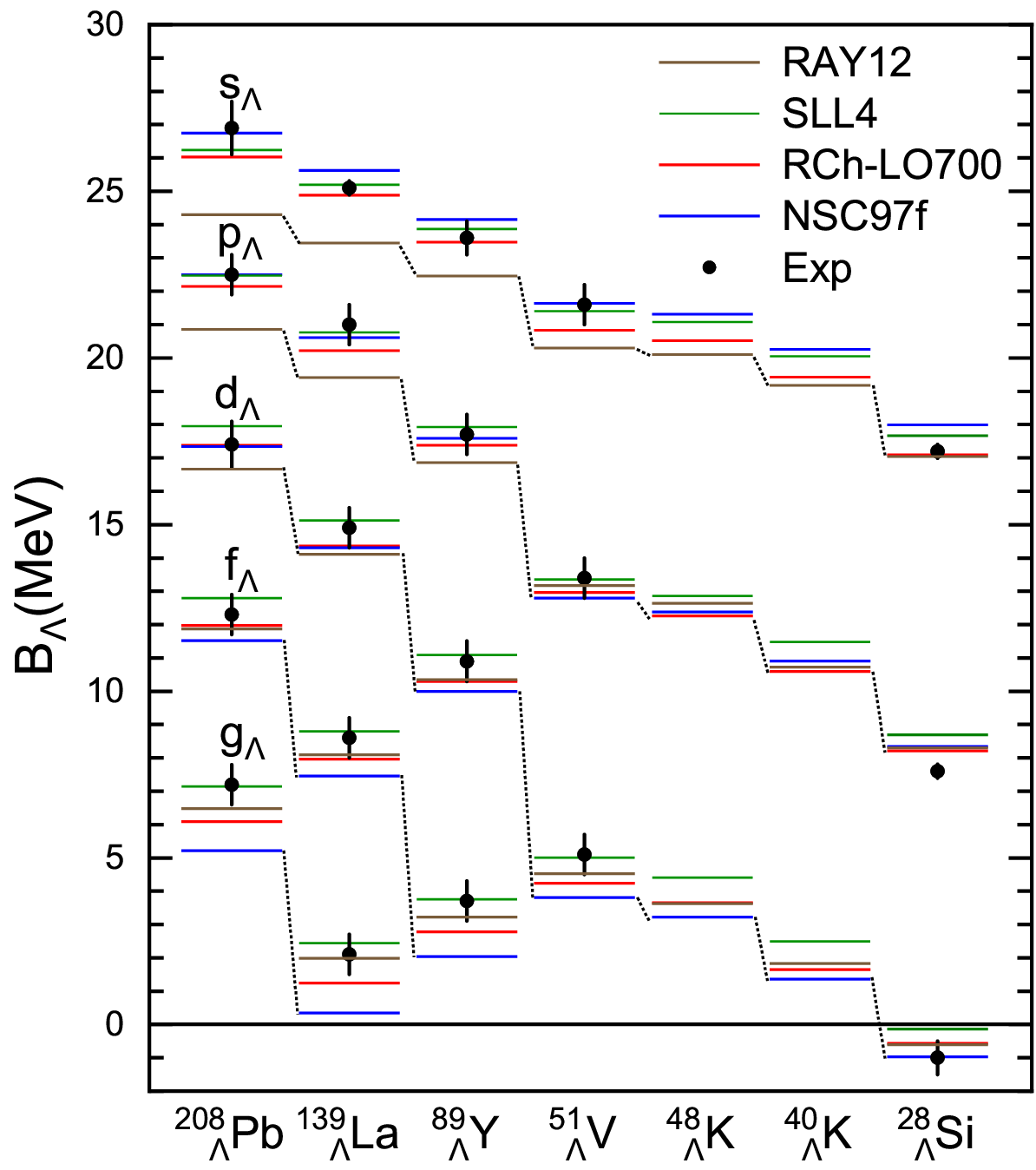}
\vspace{-2mm}
\caption{
The calculated $\la$ binding energies in different orbitals
of spherical \lpb\ and \lv\
and of \lla, \lys, \lka, \lks, \lsi\
at their deformation minima,
obtained with different interactions.
The experimental data are from Ref.~\cite{Gal16}.
}
\label{f:BPb}
\end{figure}

Since orbital-dependent $B_\la$ values are the observables most directly
related to the present analysis,
we first examine their predictions using the different interactions
for a variety of hypernuclei.
As shown in Fig.~\ref{f:BPb},
all four interactions are generally consistent with the available data
within the experimental uncertainties \cite{Gal16}.
While the SLL4 force has been globally fitted to all available
current hypernuclear data and therefore provides the best overall fit,
the main deviations are the systematic underbinding of RAY12 results
in the heavy-mass region
and the slightly too weak binding predicted by RCh-LO700 and NSC97f
for the $g$ orbital in heavy hypernuclei and the $d$ orbital in \lv.

SLL4 is parameterized directly from experimental data
\cite{Schulze14,Schulze19}.
As a result, its phenomenological nature ensures that its predictions
are generally closer to the experimental uncertainties,
making it a highly successful set of phenomenological parameters.
RAY12 is also a phenomenological interaction,
but it was mainly fitted to some light hypernuclei \cite{Rayet81}.
Consequently, it exhibits larger deviations for heavy nuclei,
especially in neutron-rich systems,
compared to the other interactions.
Its behavior shows a distinctly different trend
between lower and higher orbitals.
It slightly underbinds lower orbitals,
but higher orbitals are better reproduced.
This trend is opposite to that of NSC97f.
Among all interactions,
NSC97f exhibits the strongest binding in the lower orbitals,
but becomes weakly binding in the higher orbitals.
This effect is correlated with the effective mass of the interactions,
the largest (smallest) for RAY12 (NSC97f),
as shown in Fig.~\ref{f:meff}.

Future high-resolution experimental results of the orbital-dependent
$\la$ binding energies in \lpb\ from J-PARC \cite{Sakuma22}
will be essential to fine tune the theoretical forces
and improve the understanding of their microscopic origin.
The stronger binding of \lka\ than \lks\ is due to the larger core,
since neither interaction currently includes isospin effects.
Once experimental observations become available,
investigating isospin effects
for refining and extending the existing interactions
will become possible \cite{Nakamura22,Gogami24}.

\subsection{Environmental density
}

In contrast to the relatively light hypernuclei \lks\ and \lka,
heavy hypernuclei like \lpb\ provide systems
in which the $\la$ can be bound in higher orbitals up to the $g$ states.
This is the motivation for J-PARC HIHR to investigate
the density dependence of the $\la N$ interaction
through high-resolution measurements of the binding energies of \lpb\
\cite{Sakuma22}.
Therefore, we aim to introduce a quantity to characterize the core density
in the region occupied by the $\la$ in different orbitals.
We define the overlap integral between the $\la$ and the core density
as a measure of the average nuclear density experienced by the $\la$
in different orbitals,
and refer to it as the environmental density,
\be
 \renv \equiv \frac{\int d^3\rv\; \rol(\rv)\ron(\rv)}
 {\int d^3\rv\; \rol(\rv)} \:.
\label{e:env}
\ee

\begin{figure}[t]
\vspace{-1mm}
\includegraphics[scale=0.44]{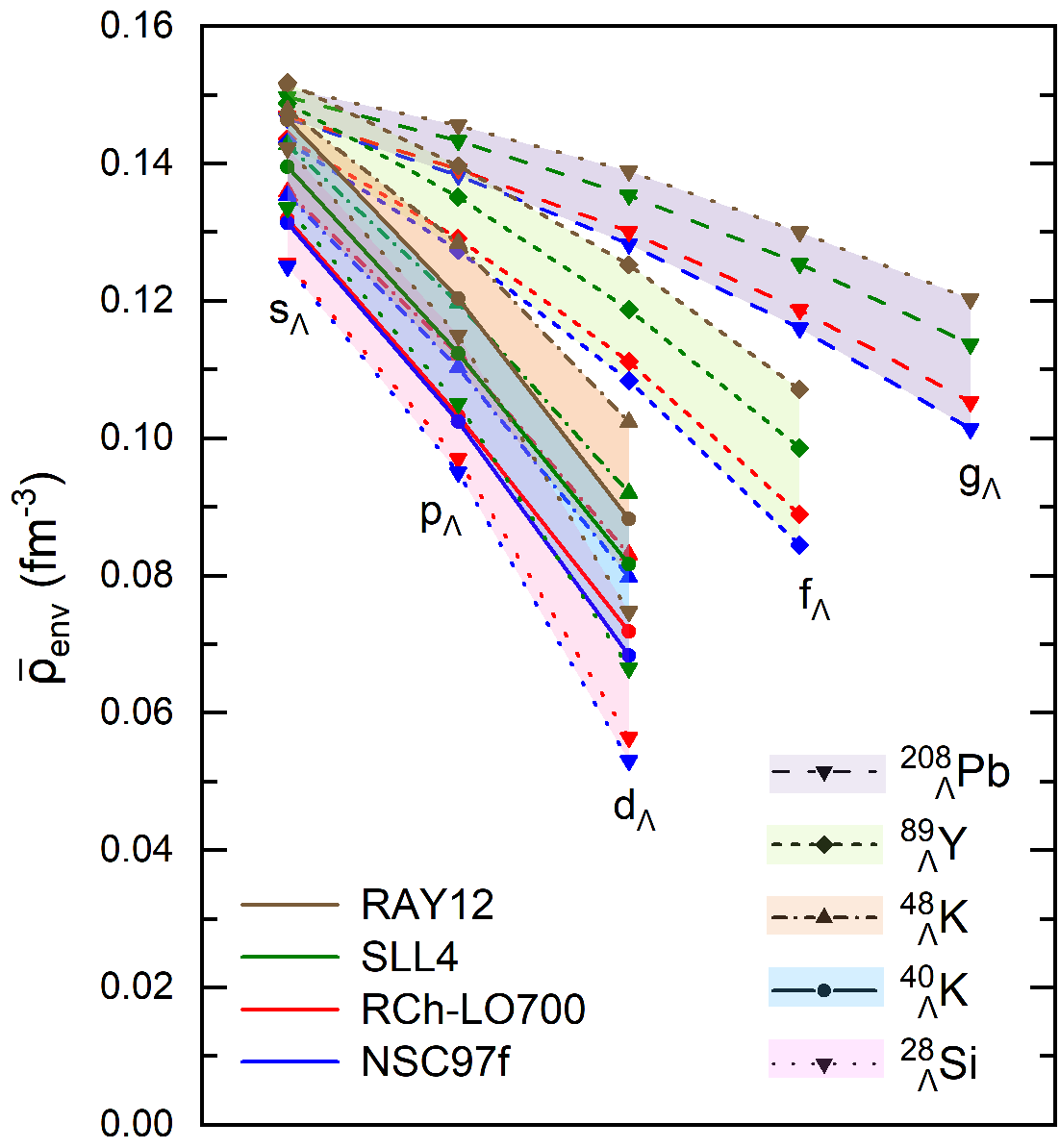}
\vspace{-1mm}
\caption{
The environmental densities,
Eq.~(\ref{e:env}),
for the $\la$ occupying different orbitals in \lpb, \lys, \lka, \lks, and \lsi,
calculated with different interactions.
}
\label{f:envPb}
\end{figure}

\begin{figure}[t]
\vspace{-1mm}
\includegraphics[scale=0.44]{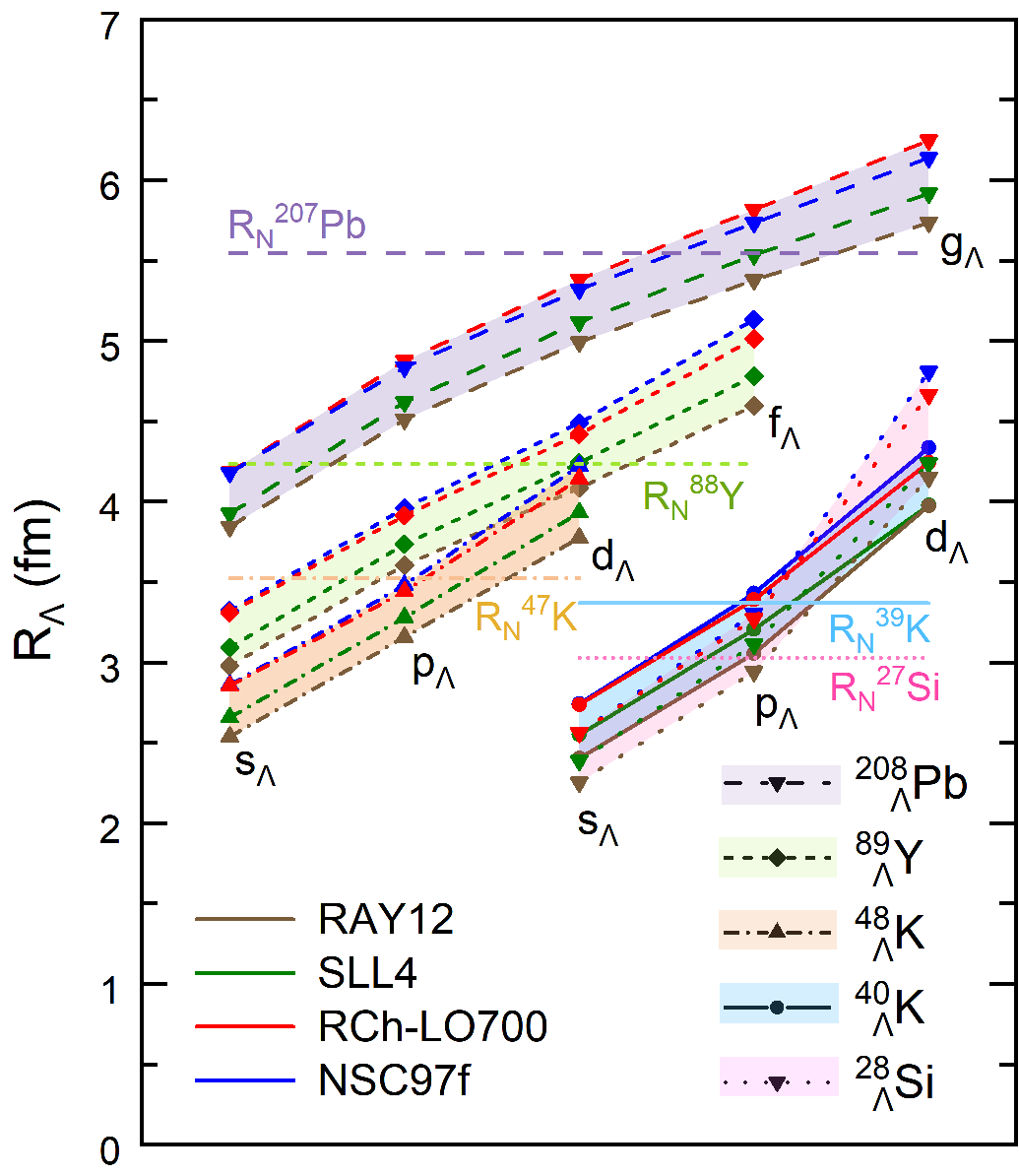}
\vspace{-1mm}
\caption{
The hyperonic radius for the $\la$ hyperon occupying different orbitals
in \lpb, \lys, \lka, \lks, and \lsi,
calculated with different interactions.
The horizontal lines indicate the various core radii.
}
\label{f:RL}
\end{figure}

Fig.~\ref{f:envPb} shows the environmental densities
of the available orbitals in
\lpb, \lys, \lka, \lks, and \lsi\ calculated using the different interactions.
The four interactions exhibit a systematic trend, namely
for the same orbital, RAY12 gives the largest environmental density,
followed by SLL4, RCh-LO700, and NSC97f in order.
The microscopic interactions show thus a tendency for the $\la$ hyperon
to move farther away from the nuclear center in higher orbitals
and to experience lower environmental densities
compared to the RAY12 and SLL4 interactions.
This correlates with smaller binding of the higher orbitals,
as shown in Fig.~\ref{f:BPb},
and with smaller effective masses, as shown in Fig.~\ref{f:meff}.

The effective mass affects both the kinetic energy
and the spatial extension of the $\la$ wave function,
which in turn influences the environmental density.
RAY12 with an effective mass of unity gives the smallest variation in $\renv$,
consistent with the smaller kinetic-energy penalty
for a localized $\la$ distribution
within the high-density nuclear core.
In contrast, NSC97f, which has the smallest effective mass,
tends to acquire larger kinetic energy
and thus moves away from the high-density core region
\cite{Vidana01,Zheng25,Zeng26}.

To confirm this mechanism,
we show in Fig.~\ref{f:RL}
the mean hyperonic radii
$R_\la = \sqrt{\int d^3\rv \rol(\rv) |\rv|^2}$
for the same set of hypernuclei as in Fig.~\ref{f:envPb}.
It can be seen that indeed the radii
exhibit a trend consistent with the environmental density,
namely for the same orbital in a given hypernucleus,
RAY12 predicts the smallest hyperon radius,
followed by SLL4, RCh-LO700, and NSC97f.
The figure shows also the nuclear core radii $R_N$ for comparison.
As expected, the low $\la$ orbitals reside well inside the core,
whereas the highest orbitals are pushed outside.

\subsection{Correlations}

\begin{figure*}[t]
\vspace{-0mm}
\includegraphics[scale=0.42]{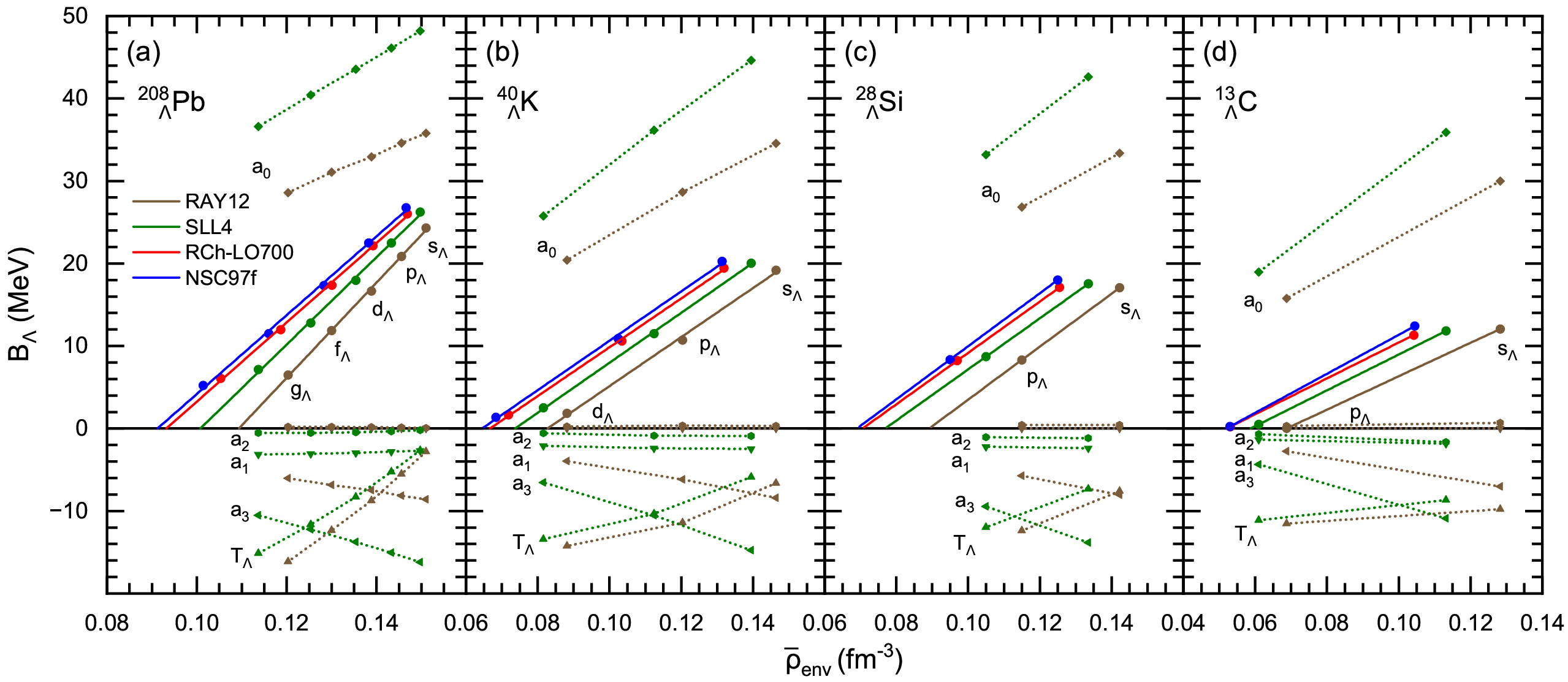}
\vspace{-6mm}
\caption{
The relation between the $\la$ binding energy
and the corresponding environmental density
in different orbitals
for various hypernuclei,
calculated using different interactions.
Markers represent the calculated values of $\renv$ and $B_\la$,
and the straight solid lines show fits $B_\la = -T_0 + A_0\renv$.
For the phenomenological Skyrme interactions,
we also show the separate contributions to $B_\la$
from $T_\la$ and the $a_1$, $a_2$, $a_3$ terms
in different orbitals (dotted curves).
Here, $T_\la$ is evaluated without effective-mass corrections
to avoid double counting with the $a_1$ term.
}
\label{f:Brho}
\end{figure*}

Having established that $\renv$
characterizes the orbital-dependent spatial environment of the $\la$,
we next examine how it enters $B_\la$ through the SHF energy-density functional.

The calculated results
for \lpb\ (a), \lks\ (b), \lsi\ (c), and \lcl\ (d)
in Fig.~\ref{f:Brho} exhibit an approximately linear
relation between $B_\la$ and $\renv$
for the different bound orbitals of a given hypernucleus and interaction.
(Note however that the results for the light nuclei \lsi\ and \lcl\
with only $s$ and $p$ bound states
are linear by construction.)
For the phenomenological interactions,
by definition $\renv=E_0/a_0$,
where $E_0$ is the contribution of the $a_0$ term to the total energy,
Eq.~(\ref{e:sll}).
Similarly, for the microscopic interactions, $\renv=-E_1/\eps_1$,
Eq.~(\ref{e:nsc}).
Thus, the leading interaction contribution is exactly linear in $\renv$
for both classes of interactions.
For the phenomenological interactions, this gives
$B_\la \approx -T_\la - a_0\renv + C_\la$,
where $T_\la$ is the kinetic energy
and $C_\la$ collects the higher-order contributions.

To understand how these additional contributions
affect the leading linear behavior,
we examine SLL4 and RAY12,
for which the individual terms in the energy-density
functional have transparent physical interpretations.
Their individual contributions to $B_\la$ are also shown in Fig.~\ref{f:Brho}.
The $a_2$ surface contribution is relatively small in all cases.
The $a_1$, $a_3$, and the kinetic contribution $-T_\la$ exhibit
repulsive contributions,
compensating the dominant attractive $a_0$ term,
but also depending approximately linearly on $\renv$,
in particular for \lpb.
Furthermore, also the density dependencies of the $a_3$ and $T_\la$ terms
tend to compensate each other:
As the $\la$ occupies a higher orbital and extends toward the nuclear surface,
its overlap with the high-density interior decreases,
reducing the repulsive $a_3$ contribution,
while the nodal and centrifugal motion increases the kinetic-energy cost.
Since both effects are governed by the same self-consistent wave function,
their variations are correlated and partly compensate each other.
Nuclear-density saturation and the limited density range
covered by the bound orbitals further suppress the remaining curvature.

Consequently, the combined contributions can be approximated
by a constant term plus a linear correction over the range considered here,
$B_\la = -T_0 + A_0\renv$,
as shown by the solid lines in the figure.
Here, $T_0$ contains the effective constant part
of the kinetic and higher-order contributions,
whereas $A_0$ contains both the leading $a_0$ contribution and the remaining
linear dependence of the additional terms.
Thus, $A_0$ and $T_0$ summarize the combined effects of the different terms
and should not be identified directly with individual interaction parameters.
The nonlinear contributions that cannot be represented by these two parameters
give rise to very small residual deviations from the fitted linear relation.

For \lpb, the fitted parameters are
$T_0=63.1$, 53.4, 44.8, $43.6\mev$, and
$A_0=577$, 530, 481, $478\pfmq$
for RAY12, SLL4, RCh-LO700, NSC97f, respectively,
ordered in the same way as the $\renv$ and $R_\la$ result
in Figs.~\ref{f:envPb} and \ref{f:RL}.
The fitted $A_0$ values are substantially different
from the corresponding leading interaction coefficients,
$-a_0=237$ and $322\pfmq$ for RAY12 and SLL4 and
$\eps_1=369$ and $384\pfmq$ for RCh-LO700 and NSC97f.
This difference shows that the kinetic and higher-order contributions
modify not only the intercept but also the slope of the fitted relation.
In particular, $T_0$
should be regarded as an effective extrapolated intercept
rather than a kinetic energy only.

For \lks, fewer bound orbital groups are available than for \lpb,
and the fit is therefore based on only three independent points.
We obtain
$T_0=24.6$, 22.3, 19.7, $19.3\mev$, and
$A_0=297$, 303, 296, $299\pfmq$
for RAY12, SLL4, RCh-LO700, and NSC97f, respectively.
Both parameters are smaller than those obtained for \lpb,
due to the smaller system.
Since the $A_0$ values are very similar
and the extrapolation is constrained by fewer orbitals,
their detailed ordering and the threshold values
should be interpreted with caution.

\OFF{
Within the linear approximation
$B_\la = -T_0+A_0\renv$,
the zero-binding intercept defines an effective threshold environmental density,
$\renv^\text{thr} = T_0/A_0$,
above which $B_\la>0$.
For \lpb, the extrapolated threshold values are
0.109, 0.101, 0.093, and $0.091\fmq$
for RAY12, SLL4, RCh-LO700, and NSC97f, respectively.
These values characterize the fitted $B_\la$--$\renv$ relations
and should not be interpreted as exact physical binding boundaries
outside the calculated density range.
}

\begin{figure}[t]
\vspace{0mm}
\includegraphics[scale=0.42]{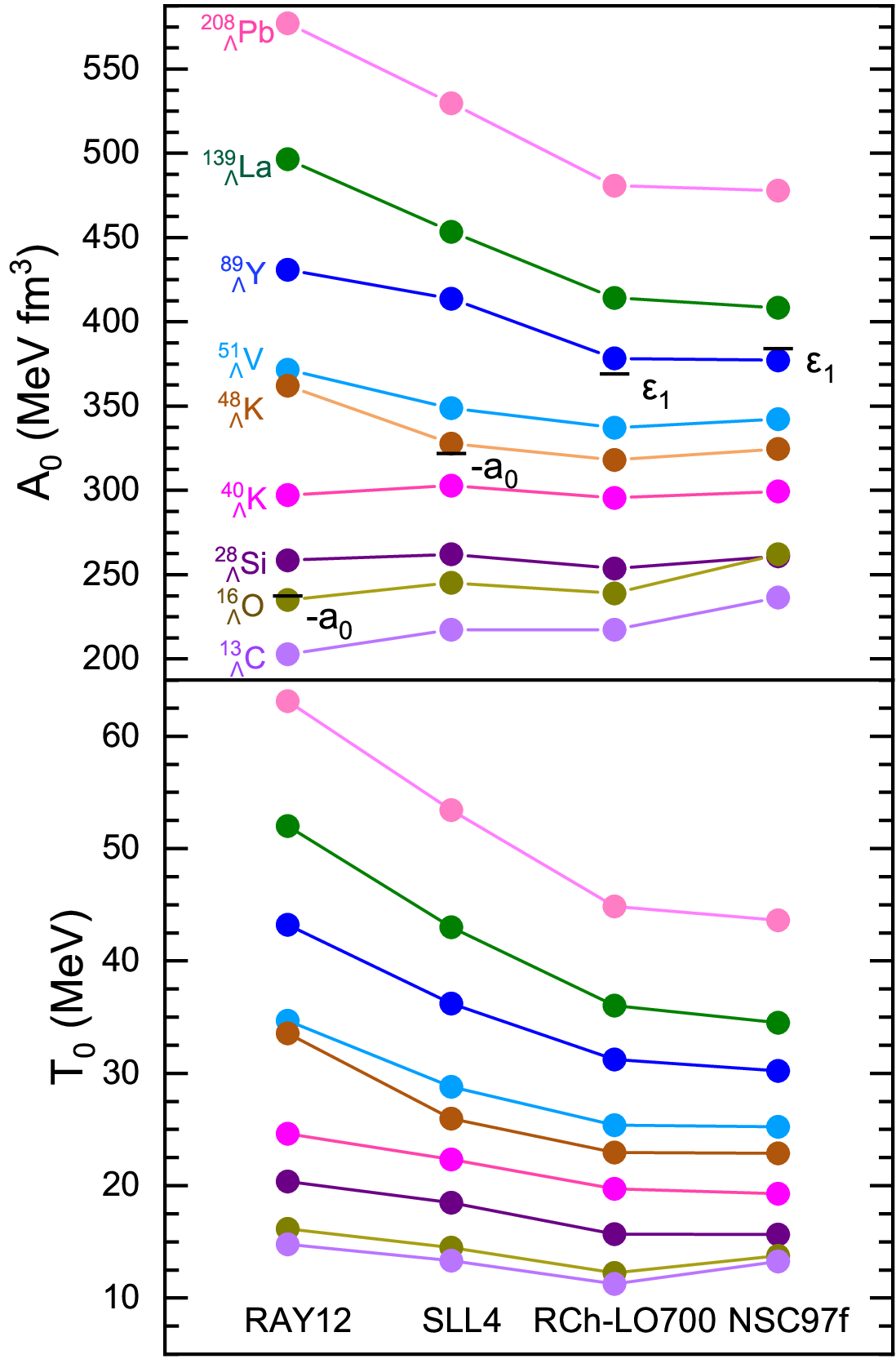}
\vspace{-0mm}
\caption{
The fitted parameters $T_0$ and $A_0$ of the relation $B_\la=-T_0+A_0\renv$
for several hypernuclei and different interactions.
The values of $-a_0$ and $\epsilon_1$ of the different forces are also indicated.
}
\label{f:all}
\end{figure}

In Fig.~\ref{f:Brho} it can be seen that
with decreasing system size,
the density dependence of the kinetic $T_\la$ contribution becomes weaker,
and this leads to smaller $A_0$ parameters closer to $-a_0$.
This becomes clear in Fig.~\ref{f:all},
where we present the values of the effective parameters $T_0$ and $A_0$
for a wide range of hypernuclei,
\lpb, \lla, \lys, \lv, \lka, \lks, \lsi, \lo, and \lcl\
using the four interactions.
The relevant values of $-a_0$ and $\epsilon_1$
for the different interactions are also indicated.

The extracted $T_0$ and $A_0$ values generally increase with mass number.
This trend reflects an effective finite-system behavior
rather than a direct change of a single interaction coefficient.
As the mass number increases,
the $\la$ binding energy, potential depth,
and number of available bound orbitals all increase as well,
whereas the environmental density of the $s$ state varies comparatively little.
Consequently, $T_0$ and $A_0$ incorporate these mass-dependent effects.

For most medium-heavy nuclei,
both $T_0$ and $A_0$ are ordered as the $\la$ effective mass:
RAY12, SLL4, RCh-LO700, and NSC97f.
This ordering can be understood from
$A_0 \simeq \Delta B_\la/\Delta\renv$:
RAY12 keeps the $\la$ distribution relatively compact
due to the large effective mass,
and produces a smaller $\Delta\renv$ across orbitals,
leading to a larger $A_0$ when $\Delta B_\la$ remains comparable.
This is related to a large positive slope of the $T_\la(\renv)$
relation in Fig.~\ref{f:Brho}.

In the lightest systems considered ($A<28$),
the ordering is reversed,
with NSC97f giving the largest $A_0$ and RAY12 the smallest.
This is due to the reduced density dependence of the $T_\la$ contribution,
which means that $A_0$ becomes uncorrelated with the effective mass,
but rather with the $a_0$ and $a_3$ parameters.
Therefore, a correlation between $A_0$ and $m_\la^*$
persists only in heavy nuclei
with a large variation of the $T_\la$ contribution between orbitals.

\subsection{Improving effective $\pmb{\la N}$ interactions}

The present analysis suggests a practical strategy for improving effective
$\la N$ interactions in finite hypernuclei.
The fitted $A_0$ and $T_0$ values provide compact finite-system descriptors
of the relation between orbital-dependent $B_\la$ values and $\renv$,
while incorporating higher-order density-dependent,
gradient, kinetic, and effective-mass effects.

For the recently developed RCh-LO700,
which already provides an overall good description of $B_\la$
across a broad mass range \cite{Zeng26},
the present results extend its validation to the effective density dependence
in finite hypernuclei.
Future measurements of orbital-dependent $B_\la$ values can
then guide further refinement of RCh-LO700.
SLL4 provides a useful phenomenological reference,
while corresponding measurements can further constrain the balance between
its leading volume, nonlinear density-dependent, and surface contributions,
as well as control the introduction of additional terms
in the energy-density functional.
RAY12 was constructed only for light hypernuclei
and therefore underbinds heavy ones.
NSC97f instead is a completely theoretical force
without free parameters
based on a $\la N$ potential
and not fitted to any hypernuclear data;
it does not and cannot be expected to
yield results of the same quality as the other interactions.

The planned measurements of \lpb, \lks, and \lka\
\cite{Sakuma22,Nakamura22,Gogami24}
will therefore provide complementary constraints on these interactions.
In particular, the several bound orbitals in \lpb\
cover a relatively broad range of environmental densities,
whereas \lks\ and \lka\ test lighter and more isospin-asymmetric systems.
Because fewer bound orbitals are available in the latter nuclei,
their results should be combined with the heavy-hypernuclear data
in a global refinement.

\section{Summary}
\label{s:conc}

Given the limited direct constraints from $YN$ scattering experiments
and femtoscopy,
finite hypernuclei provide an important testing ground for effective
$\la N$ interactions.
Their orbital-dependent $B_\la$ values are directly accessible observables
sensitive to the effective density dependence.
Relating these observables to that density dependence requires a common
quantitative measure of the nuclear environment experienced by the $\la$
in each orbital,
together with compact representative parameters for
comparing different interactions.
Such an analysis is also particularly relevant for further examining
the recently developed microscopic interaction RCh-LO700 in finite hypernuclei.
We introduce the environmental density $\renv$,
defined as the overlap integral of the $\la$ density and the core density,
to quantify the nuclear environment
associated with orbital-dependent $\la$ binding energies.
Evaluating this quantity for
\lpb, \lka, \lks, and several other representative hypernuclei,
we systematically compare the density-dependent and effective-mass
characteristics of the phenomenological interactions RAY12 and
SLL4 and the microscopic interactions RCh-LO700 and NSC97f.

For a given hypernucleus and a given $\la N$ interaction,
the $\la$ binding energies in different orbitals
exhibit an approximately linear dependence on the environmental density,
$B_\la=-T_0+A_0\renv$.
The parameters $A_0$ and $T_0$ offer a compact characterization
of this dependence in finite hypernuclear systems.
By associating the orbital-dependent $B_\la$ values with their corresponding
environmental densities,
hypernuclear spectroscopy can constrain the effective density dependence
of the $\la N$ interaction in finite systems.
The approximate linearity is not a prerequisite for such constraints,
but allows them to be summarized compactly by the two effective parameters
$A_0$ and $T_0$.

For \lpb, the environmental density decreases
from about $0.15\fmq$ for the $s$ state
to about $0.10\fmq$ for the $g$ state,
reflecting the increasingly extended distribution
of the $\la$ hyperon in higher orbitals.
For \lks\ and \lka,
the larger core of \lka\ leads to stronger $\la$ binding than in \lks.
Since the interactions considered here do not include explicit isospin effects,
future data for these systems will be useful for examining the missing
isospin dependence.

The fitted $A_0$ and $T_0$ values
generally increase from light to heavy hypernuclei,
although their ordering among the interactions depends on the
mass region and the available bound orbitals.
We found a significant correlation with the effective $\la$ mass
of the interaction for heavy nuclei:
A larger effective mass favors larger environmental densities
and therefore larger $T_0$ and $A_0$.
Together with the absolute $B_\la$ values,
these systematics provide interaction-specific guidance,
in particular, the orbital-dependent behavior
of the recently developed RCh-LO700 should be validated
using both the absolute $B_\la$ values
and the extracted $A_0$ and $T_0$ systematics.
The present framework therefore provides quantitative finite-system constraints
for testing and refining effective $\la N$ interactions.

\vspace{10mm}
\medskip
\section*{Acknowledgments}

This work was supported by the National Natural Science Foundation of China
under Grant Nos.~12575124, 12688101, and 12547102,
and by the Fundamental Research Funds for the Central Universities.


\newcommand{\epja}{EPJA}
\newcommand{\npa}{Nucl. Phys. A}
\newcommand{\nphysa}{Nucl. Phys. A}
\newcommand{\ppnp}{Prog. Part. Nucl. Phys.}
\newcommand{\ptp}{Prog. Theor. Phys.}
\newcommand{\ptep}{Prog. Theor. Exp. Phys.}
\bibliography{ddep}
\end{document}